\documentclass[aps,
               prl,
               amsmath,
               amssymb,
               preprint,
               reprint,
               showkeys,
               showpacs]
               {revtex4-1}

\usepackage[T1]{fontenc}
\usepackage{dcolumn}
\usepackage{bm}
\usepackage{graphicx}
\usepackage{multirow}
\usepackage{hyperref}
\usepackage{color}

\newcommand{\CP}{CePt$_5$}
\newcommand{\LP}{LaPt$_5$}
\newcommand{\tn}{$t_{\text{nom}}$}

\begin{document}

\begin{abstract}
Low-temperature electronic properties of rare-earth intermetallics are substantially influenced by the symmetry and magnitude of the crystal electric field. 
The direct spectroscopic analysis of crystal field splitting can be challenging, especially in low-dimensional systems, because it requires both high spectral resolution and pronounced sensitivity. We demonstrate the eligibility of electronic Raman spectroscopy for this purpose by the direct determination of the $4f$ level splitting in ultra-thin ordered {\CP} films  down to 1.7 nm thickness on Pt(111). Crystal field excitations of Ce $4f$ electrons give rise to Raman peaks with energy shifts up to $\approx$ 25~meV.
Three distinct peaks occur which we attribute to inequivalent Ce sites, located (i) at the interface to the substrate, (ii) next  to the Pt-terminated surface, and (iii) in the {\CP} layers in between. The well-resolved Raman signatures 
allow us to identify a reduced crystal field splitting at the interface and an enhancement at the surface, highlighting its strong dependence on the local atomic environment. 
\end{abstract}

\date{\today}

\title{Study of crystal-field splitting in ultrathin {\CP} layers by Raman spectroscopy}

\author{B. Halbig}
\email[email: ]{Benedikt.Halbig@physik.uni-wuerzburg.de}
\thanks{Both authors contributed equally to this work}
\author{U. Bass}
\author{J. Geurts}
\affiliation{Universit{\"{a}}t W{\"{u}}rzburg, Physikalisches Institut, Experimental Physics 3, Am Hubland, 97074 W{\"{u}}rzburg, Germany}

\author{M. Zinner}
\thanks{Both authors contributed equally to this work}
\affiliation{Universit{\"{a}}t W{\"{u}}rzburg, Physikalisches Institut, Experimental Physics 2, Am Hubland, 97074 W{\"{u}}rzburg, Germany}
\author{K. Fauth}
\affiliation{Universit{\"{a}}t W{\"{u}}rzburg, Physikalisches Institut, Experimental Physics 2, Am Hubland, 97074 W{\"{u}}rzburg, Germany}
\altaffiliation{RCCM,Universit{\"{a}}t W{\"{u}}rzburg, Am Hubland, 97074 W{\"{u}}rzburg, Germany}

\maketitle

In metallic systems containing $4f$ elements, the hybridization of localized
$4f$ states with conduction electrons, combined with strong local Coulomb correlation, may lead to various manifestations of the fascinating field of Kondo and heavy fermion physics \cite{H93,GS91,LRV+07,Y16,SW16,P16}.
Thermodynamic and magnetic properties in these materials strongly depend on the $4f$ level splitting by the crystal electric field (CF).
Inelastic neutron scattering (INS), well established for the determination of this splitting in bulk materials, requires large sample volumina. The rising interest in low-dimensional correlated systems calls for complementary spectroscopic methods which are applicable to ultrathin films.

\begin{figure}[htbp]
\includegraphics[width=\columnwidth]{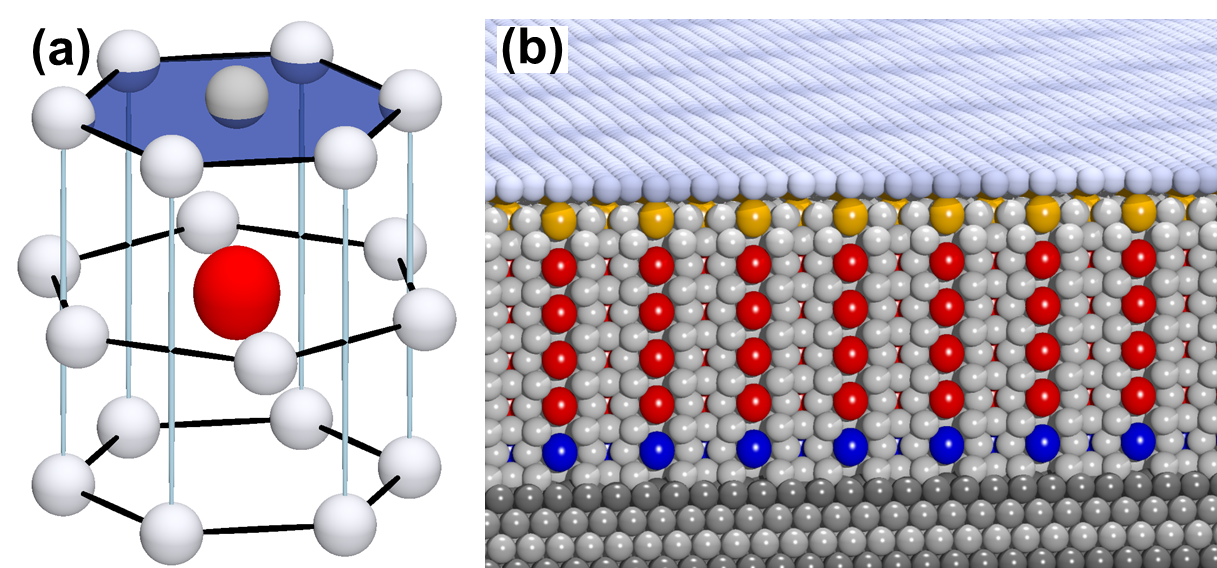} 
\caption{(Color online) {\CP} lattice structure. (a) Ce atom (large red sphere) within the hexagonal elementary cell. (b) cross section of a 6 u.c. {\CP} layer on Pt(111). Same color code for inequivalent Ce atoms (i.e. in proximity of the surface, the interface, or within the film) as in Fig.~\ref{fig:IntensitaetDickeLTCePt5} and Fig.~\ref{fig:DickenabhaengigCFE1bisCFE2}.}
\label{fig:CePt5}
\end{figure}

An interesting representative of the latter kind of systems is the interdiffusion-induced binary intermetallic Ce-Pt surface phase {\CP} with a thickness of few unit cells on a Pt(111) substrate \cite{PZK+15,KNS+11}. 
The atomic lattice of the {\CP} surface layer is based on the CaCu$_5$ structure, as shown in Fig.~\ref{fig:CePt5}(a). Its symmetry is hexagonal, consisting of alternating CePt$_2$ layers and Pt-layers. While the atoms in the Pt layers of the regular {\CP} lattice form kagome structures, in the outermost Pt monolayer the kagome hole positions are filled, resulting in a dense hexagonal Pt-terminated surface \cite{PZH+15,TPS15}.

The observation of Kondo screening in {\CP}/Pt(111) \cite{PZK+15,PF16} may appear quite remarkable since bulk {\CP} was reported to not exhibit Kondo interactions \cite{LMK+79}. 
Valuable information on the details of the underlying mechanisms and interactions on the atomic level was obtained from intrinsically surface-specific methods, such as soft x-ray absorption (XAS) and magnetic circular dichroism (XMCD) \cite{PZK+15,PF16}. 
It is non-trivial, however, to independently assess the crystal field splitting and the quantitative strength of Kondo screening which both affect the magnitude and temperature dependence of the paramagnetic response \cite{PZH+16,PF16}.
Therefore, an independent method for determining the CF-induced Ce $4f$ level splitting in such ultrathin layers is highly desirable.
 
In recent years, Raman spectroscopy (RS) as an optical technique, commonly used for investigating vibration excitations in nonmetallic bulk and multilayer systems such as semiconductor heterostructures \cite{G96}
has advanced to a sensitive probe for 
vibration eigenmodes of surfaces as well as 
ordered atomic overlayers  
\cite{E99,ER00,RSE+12,WWG+02,LBB+14}.
This development was boosted by exploiting resonance enhancement of the excitation process and by the significant improvement of the detection sensitivity. 
Notable advantages of RS are its high spectral resolution and its ability to reveal excitation symmetry properties by utilizing well-defined light polarization configurations. 
Apart from vibration mode studies, RS has also been utilized for investigating electronic excitations, including a few studies of CF excitations in rare earth materials \cite{S00}. 
To our knowledge, to this date all published Raman results from $4f$ electrons originate from bulk samples \cite{GJB+83,ZHB+84,GZB+85,CKF+86}.
  
In this letter, we report on the direct determination of the CF-induced $4f$ level splitting in ultrathin {\CP} layers (between 3.5 and 18.0 unit cells) from crystal field excitations (CE) in $in~situ$ electronic Raman scattering in UHV.
Three distinct CE Raman peaks are identified. 
Their electronic nature follows from both their temperature dependence and their absence in isostructural {\LP}  films without $4f$ electrons, whereas both materials feature very similar vibrational Raman losses.
The CE peaks of {\CP} exhibit individual evolutions of intensity vs. {\CP} film thickness.
This allows us to assign them to spatially distinct Ce positions in the specimens, i.e., at the {\CP}/Pt(111) interface, `bulk-like' sites, and those adjacent to the Pt surface termination layer.

The preparation of ordered Ce-Pt and La-Pt phases has been reported in the literature \cite{KPK+14,PZH+16,GPB+98}, 
and the specimens of the present study were prepared essentially according to these procedures.
In brief, clean Pt(111) surfaces were prepared by repeated cycles of $1$ keV Ar$^+$ ion sputtering and annealing at $T=1170$ K. 
The surface intermetallic phases were then generated by evaporating the desired amount of Ce (La) onto the substrate near ambient temperature and subsequently annealing the specimen to $T=920$ K for 5-10 min.
They are readily identified by their characteristic LEED patterns \cite{PZK+15}.
For the La-Pt system we observe a similar succession of structural phases as in CePt$_5$ \cite{KPK+14,PZH+15} which, however, evolves much faster as a function of coverage.
Nevertheless, from the reported evidence and similarities between the phases \cite{GPB+98,RRB00}, Auger electron spectroscopy, and the results below we conclude that the La-Pt phases consist of {\LP}.
In line with previous work we specify the intermetallic thickness $t_{\text{nom}}$ in nominal multiples of unit cells (u.c.) along the surface normal 
(1 u.c. $\approx 0.44$ nm).

Immediately after preparation, the specimens were transferred \textit{in situ} to the UHV optical analysis chamber  (residual pressure p < 2$\cdot$10$^{-10}$ mbar) and mounted onto a continuous-flow He cryostat.
An initial set of Raman spectra taken at room temperature (RT) was followed by a series of low-temperature measurements (LT, $T\approx 20$ K).
The samples were excited by the 2.54 eV and 2.41 eV lines of an Ar$^+$-ion laser (incident power: 100 mW).

The scattered light was collected in near-backscattering geometry by an f/3 lens system and analyzed by a single monochromator (SPEX 1000M) with CCD detector (ANDOR iDus series, quantum efficiency $\approx85\%$). 
Inserting an ultra-steep long-pass edge filter (SEMROCK RazorEdge) into the optical path
allowed  the detection of  Raman signals down to 9 meV from the laser line. The energy resolution (FWHM) was
approximately 0.4 meV.
Polarization dependent spectra were recorded with vertically polarized incident radiation and either vertically (`vv') or horizontally (`vh') polarized detection. 
Spectra taken without polarization selection will be denoted as `vu'.
Typical integration times for a single spectrum amounted to 900 s.

\begin{figure}
\includegraphics[width=\columnwidth]{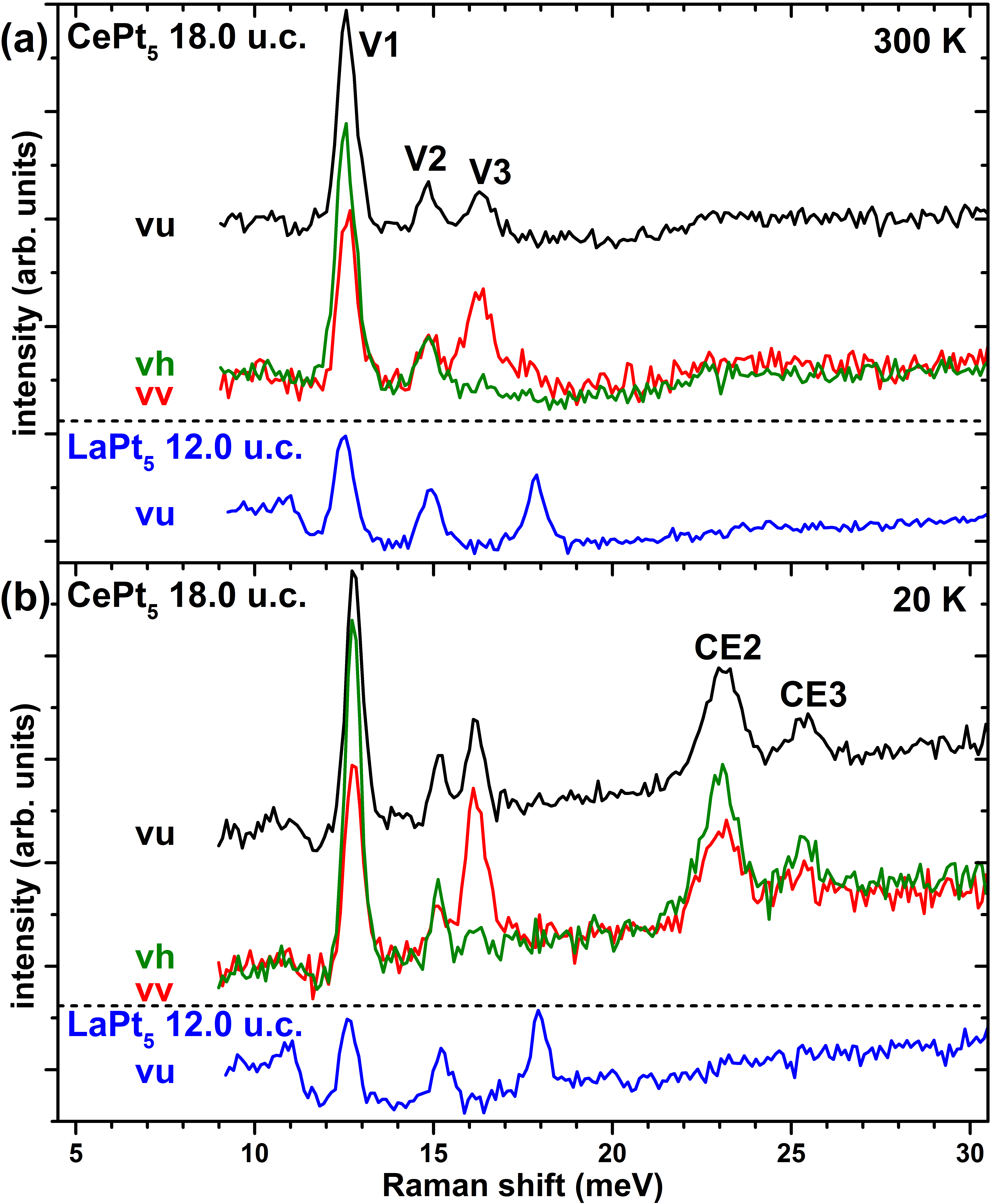}  
\caption{(Color online) (a) Raman spectra of a {\CP} surface layer (18~u.c.) on Pt(111), and a reference sample with a {\LP} surface layer (12 u.c.), taken at T = 300~K. (b) corresponding spectra at T~$\approx$~20~K. vv and vh denote the polarization directions of the incoming (v) and scattered (either v or h) light; vu: unpolarized detection. Incident photon energy: 2.54~eV.}
\label{fig:PolarisationCePtLaPtRTLT}
\end{figure}

Figure \ref{fig:PolarisationCePtLaPtRTLT}(a) displays the Raman spectra obtained at RT for the specimens of largest thickness for both {\CP} ($t_{\text{nom}}= 18$ u.c.) and {\LP} ($t_{\text{nom}}= 12$ u.c.).
For both materials the vu spectra exhibit three pronounced Raman losses between 12 and 18 meV which we label as V1, V2, and V3.
These peaks may be attributed to vibration modes owing to their polarization and thickness dependence (see Fig.~\ref{fig:IntensitaetDickeLTCePt5} below) as follows. 
Bulk {\CP} and {\LP} both crystallize in CaCu$_5$ type structure (P6/mmm, D$_{6h}^1$, No. 191) \cite{P93}.
This structure exhibits only one Raman active mode, whose symmetry is $E_{2g}$ (x$^2$-y$^2$, xy) or $\Gamma_6^+$ \cite{RBP81}.
According to its Raman tensor, this bulk mode should be visible in the vv as well as the vh configuration, as is in fact observed for peak V1. 
The assignment of feature V1 to the E$_{2g}$ mode is underscored by the dependence of its intensity on 
{\tn}, as illustrated in Fig.~\ref{fig:IntensitaetDickeLTCePt5}.
Increasing the {\CP} thickness obvioulsy entails a concomittant increase of the V1 spectral weight
which we attribute to the growing `inner volume' of the intermetallic film. 
In contrast, features V2 and V3 show much less of a thickness dependence and we therefore attribute 
them to the uppermost region of the  intermetallic film, where the lattice symmetry is reduced from D$_{6h}$ to C$_{6v}$ due to structural relaxations, as shown by a recent LEED-IV analysis \cite{PZH+15}.
As a result, four additional modes are now Raman allowed, two of which (E$_2$ and A$_1$) possess the polarization dependences observed for V2 and V3.
These assignments are readily transferred to the {\LP} specimens, since the polarization dependence of their vibration features (not shown)
is an exact match to those of {\CP}. 
At LT (Fig.~\ref{fig:PolarisationCePtLaPtRTLT}(b) the vibrational Raman features exhibit reduced line widths as well as small but systematic frequency shifts
the detailed discussion of which is deferred to a later publication \cite{BHunpub}. 

Here, we focus on the remarkable difference between the two materials instead, represented by the emergence of two new Raman features appearing in {\CP} at loss energies of (23.1 $\pm$ 0.2) meV (CE2) and (25.4 $\pm$ 0.1) meV (CE3), while obviously absent in {\LP}.
In hindsight we note that a faint contribution of these electronic transitions may already be discerned  in the RT spectra. 
A third peak will be identified further below.
No other spectral signatures were found up to 180 meV.

We attribute the additional Raman peaks in the {\CP} spectrum to Ce $4f$ CF excitations. 
This seems to be a natural choice since the main difference of Ce with respect to La consists of the extra $4f$ electron, and CF excitations generally acquire more intensity at low temperature \cite{S00}.
The appropriateness of this attribution may once again be discussed based on the polarization dependence of the new features, which is similar to that of V1.
In the CF picture the hexagonal symmetry of the Ce site in {\CP} leads to a splitting of the $j=5/2$ multiplet into three Kramers doublets of pure $m_j = \pm 1/2$, $\pm3/2$, and $\pm$5/2 character, which belong to the double group representations $\Gamma^-_7$, $\Gamma^-_9$, and $\Gamma^-_8$  \cite{S00,KP69}.
Since the quadrupole selection rules for CE Raman transition in the present geometry require $\Delta m_j = \pm2$, only the transitions involving the $m_j = \pm$1/2 doublet are symmetry allowed, i.e.  $\Gamma^-_7 \leftrightarrow \Gamma^-_9$ ($\pm1/2 \leftrightarrow \mp 3/2$) and  $\Gamma^-_7 \leftrightarrow \Gamma^-_8$ ($\pm1/2 \leftrightarrow \pm 5/2$).
Their polarization dependence is encoded in the irreducible representations contained in the direct products $\Gamma_7^- \otimes \Gamma_9^-$ and $\Gamma_7^- \otimes \Gamma_8^-$  , respectively.
Both share $\Gamma_6^+$ as a member which we already identified with the vibration mode V1.
The symmetry analysis thus leads us to expect that the CE peaks and V1 have the same polarization dependence, which is readily confirmed for CE2 and CE3 from Fig.~\ref{fig:PolarisationCePtLaPtRTLT}(b).
A closer inspection of the {\CP} Raman spectra at small {\tn} reveals that there is actually a third, broader 
excitation, as shown in the inserts to Fig.~\ref{fig:IntensitaetDickeLTCePt5}. 
With an excitation energy of (16.4 $\pm$ 0.5) meV and FWHM of $\approx 4$ meV it energetically overlaps with 
the vibration peaks V2 and V3.
Since it is absent in {\LP} and of strongly reduced intensity at RT, 
it is again associated with a CF excitation and labeled as CE1.

\begin{figure}
\includegraphics[width=\columnwidth]{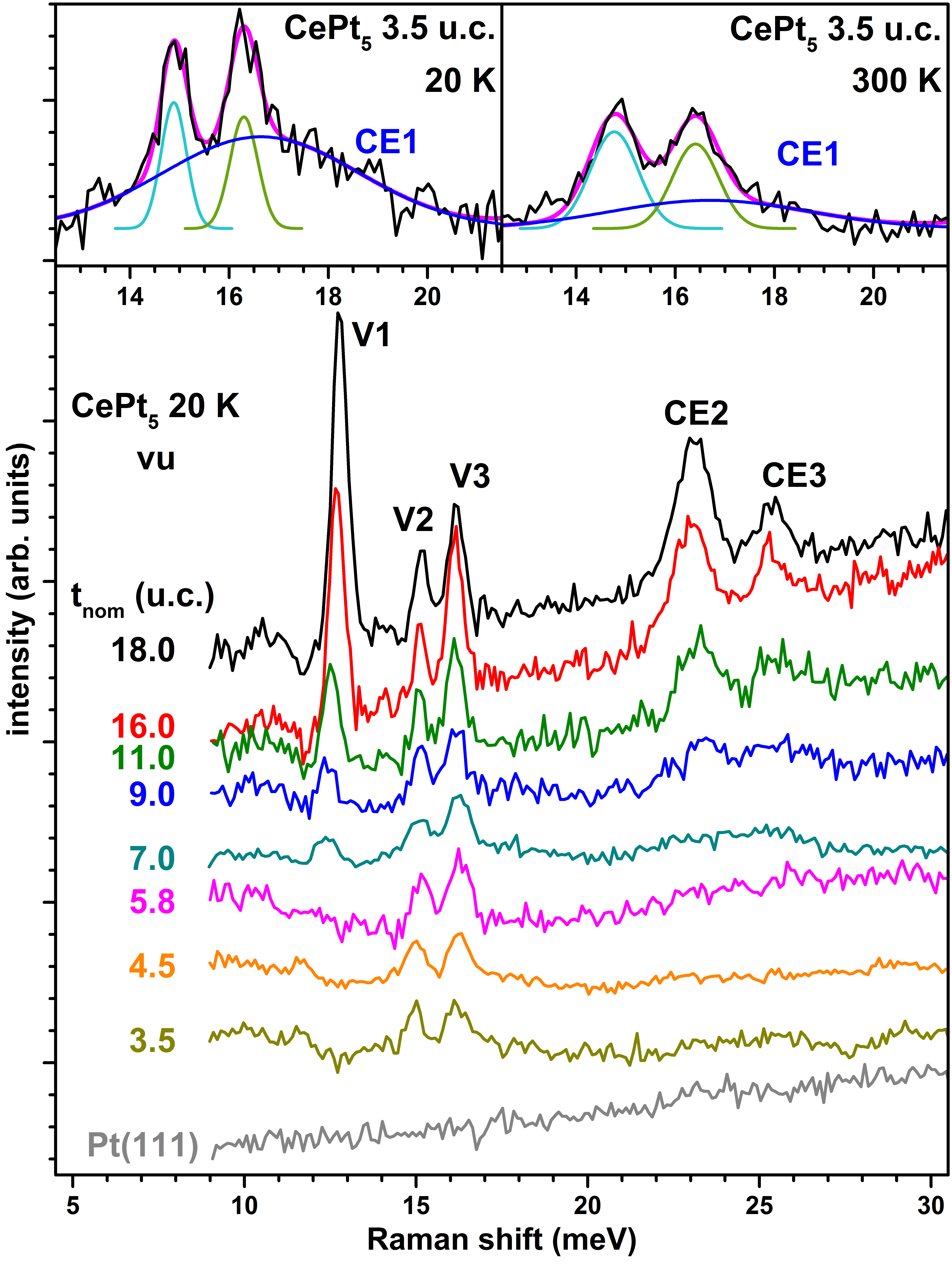} 
\caption{(Color online) Unpolarized, background subtracted Raman spectra of the bare Pt(111) substrate and {\CP} layers with thickness between 3.5 u.c. and 18 u.c., taken at $T \approx$ 20 K. Spectra  vertically offset for clarity. Insets: zooms into the 3.5 u.c. spectrum at 20 K (left) and 300 K (right), respectively, along with the fits of peaks V2, V3, and CE1.}
\label{fig:IntensitaetDickeLTCePt5}
\end{figure}

\begin{figure}
\includegraphics[width=\columnwidth]{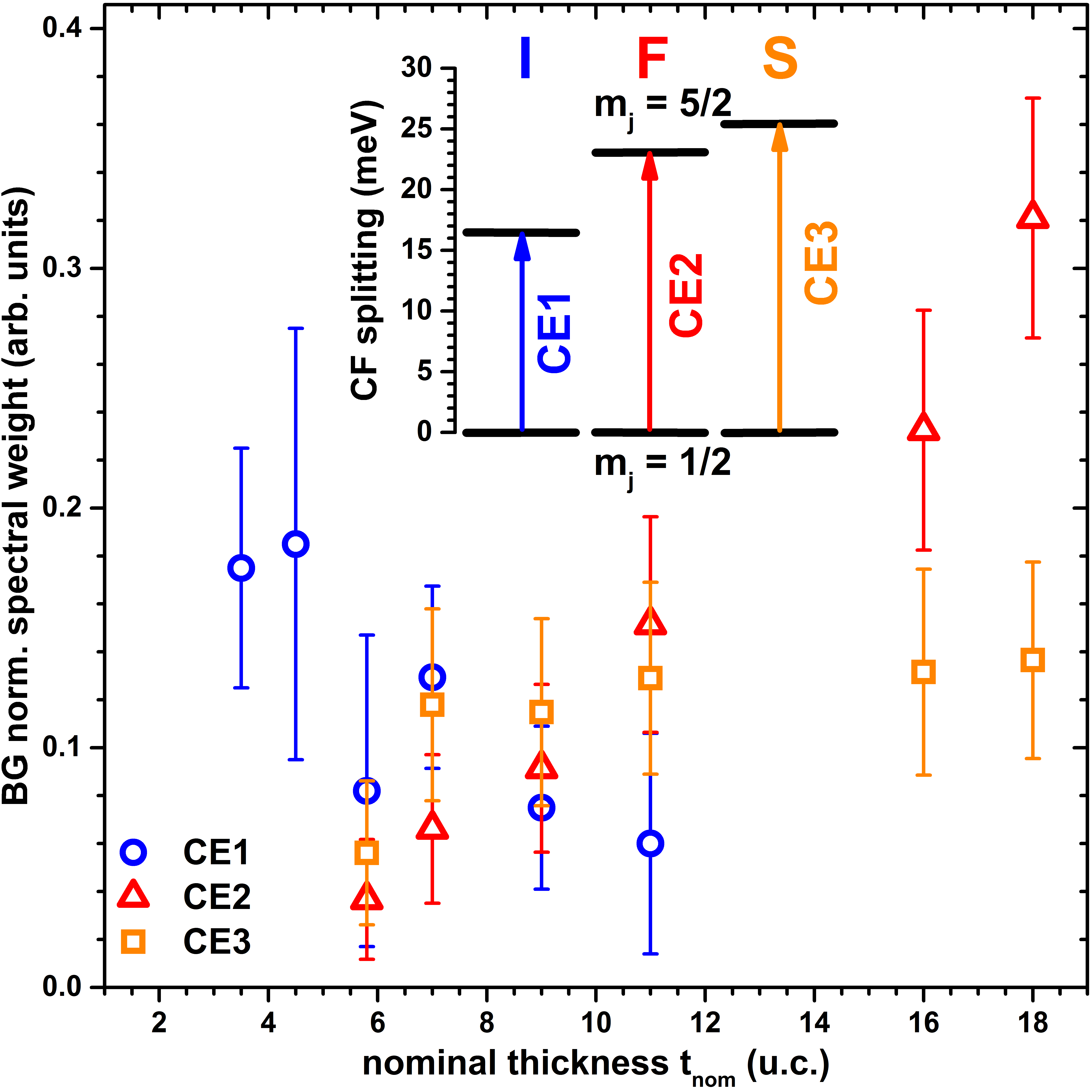}
\caption{(Color online) {\CP}-layer-thickness dependence of the Raman intensities at T = 20 K of electronic transitions CE1 to CE3 between crystal-field-split Ce 4f levels. These crystal field excitations are denoted as CE1 to CE3. Inset: schematic assignment of the CF excitations to interface (I), film (F), and surface (S).}
\label{fig:DickenabhaengigCFE1bisCFE2}
\end{figure}

The observation of three distinct CE excitations clearly cannot be reconciled with a single CF splitting scheme valid for all Ce sites alike.
The key to their understanding lies in evaluating their spectral intensities as a function of intermetallic film thickness {\tn}.
To this end, each Raman spectrum was subjected to a least squares fit by a superposition of Gaussians above a linear background.
In Fig.~\ref{fig:DickenabhaengigCFE1bisCFE2}, we report the spectral weights of the LT electronic Raman losses, normalized to the background intensity at 20 meV.
Their dependence on {\tn} is quite individual: 
while the intensity of CE1 rapidly decreases with increasing film thickness and vanishes for  {\tn}$\gtrsim 10$ u.c., the CE2 intensity shows a complementary increase, and that of CE3 is thickness-independent within experimental error for {\tn}$\gtrsim 7$ u.c.
Therefore we assign the different CF excitations to Ce atoms with different local environments in the intermetallic films.

The rapid decrease of the CE1 intensity suggests that it is related to the Ce sites in the vicinity of the interface to the Pt(111) substrate. 
Its rate of decay with {\tn} leads to an estimated attenuation length of $\lambda \approx$ (6.0 $\pm$ 1.5) nm.
The steady growth of the CE2 intensity is very reminiscent of the behavior of the vibration peak V1 of the {\CP} film and hence suggestive of associating it with the CF excitations in the `inner', bulk-like volume of the film.
Based on the $\lambda$ estimate, however, one would expect a signature of saturating intensity of CE2 as well as V1 in the limit of largest {\tn}, which is clearly not the case.
A non-saturating growth of V1  may arise from an increase of scattering efficiency upon gradually acquiring bulk-like symmetry with increasing film thickness. 
This will also lead to stronger electronic scattering provided there is significant coupling to the vibration mode.
Such coupling was indeed discussed and detected in earlier work \cite{GJB+83,LWK+02,SWl+03}.
Finally, the nearly thickness-independent intensity of feature CE3 is very compatible with CF excitations limited to the vicinity of the film surface.
The spatial dependence of the CF energies highlights that the CF splitting may sensitively depend on the local surroundings `even in a metal'.
A similar conclusion was recently reached concerning the nature of the electronic ground state in a Ce-Pd surface alloy on the basis of photoemision experiments \cite{MSJ+14}.

A conclusion with respect to the character of the CE transitions ($\Gamma^-_7 \leftrightarrow \Gamma^-_9$~vs.~ $\Gamma^-_7 \leftrightarrow \Gamma^-_8$) can be reached by using complementary information.
From XAS experiments it was concluded that the states with $m_j=\pm5/2$ are highest in energy and that the splitting between the states with $m_j=\pm1/2$  and  $m_j=\pm3/2$ amounts to less than 1 meV \cite{P15,PF16}.
On this basis we can unanimously state that the Raman losses observed here all are of $\Gamma^-_7 \leftrightarrow \Gamma^-_8$ character and represent excitations from lower-lying states with $m_j = \pm 1/2$ to excited states with $m_j = \pm 5/2$.
The temperature dependence of x-ray linear and cirular dichroism at the Ce M$_{4,5}$ edges furthermore suggested an energy scale for the total CF level spread of $\lesssim 18$ meV for {\tn} $\lesssim 4$ u.c. and $\gtrsim 25$ meV for {\tn} $\gtrsim 7$ u.c., respectively.
Given that the XAS probing depth is smaller by roughly a factor 4 to 6 compared to the present work, this level of almost quantitative agreement is extremely gratifying. 
Both sets of experiments also agree in that no higher lying CF excitations are observed in the ultrathin  {\CP} films.
Detecting the CF excitations by Raman scattering offers the compelling advantage of high spectral resolution, though.
It allows us to discriminate between coexisting Ce $4f$ schemes involving minute differences at the quantitative level which would otherwise go unnoticed.
Also, the level of accuracy achievably by direct observation of the excitations exceeds by far what can be obtained from fitting Boltzmann distributions in temperature dependent experiments.

The high spectral resolution also allows us to examine the width of the CE Raman losses and to address the question of the enhanced width of CE1 compared to CE2 and CE3.
Here we can draw an analogy with INS.
Recently, Willers et al.~\cite{WHH+10} have noted for Ce compounds a positive correlation between the width of CF excitations in INS and the strength of hybridization between Ce $4f$ electrons and itinerant states seen by XAS, suggesting that hybridization contributes significantly to reducing the lifetime of the CF excited states. 
In this respect our Raman results are fully in line with these INS observations. 
The CE1 excitation relates to the CF splitting scheme which is predominant at small film thickness and is characterized by a large width of the Raman loss peak.
This is also the thickness regime of strongest Ce $4f$ hybridization \cite{PZK+15}.
Interestingly, the regime of stronger hybridization and hence stronger Kondo screening \cite{PZK+15,PF16} correlates with the occurrence of a smaller overall CF splitting.
A similar conclusion seems to hold in the case of CeAg$_x$ films, albeit on a much reduced energy scale \cite{MZunpub}.

In summary, we have shown that besides the vibration modes also the crystal field splitting of $4f$ electron levels can be determined with high accuracy by direct observation of Raman scattering from electronic transitions between CF-split levels, even for ultrathin intermetallic films comprising just a few atomic layers and a moderate density of rare earth atoms. 
In our exemplary study of {\CP}/Pt(111)  the observation of three distinct CE Raman peaks for Ce atoms at the {\CP}-Pt(111) interface, within `bulk-like' {\CP} layers, and  adjacent to the film surface, reveals the different crystal fields of these atomic environments.
The small energy difference between CE2 and CE3 in particular could hardly have been resolved by other means. 
The electronic origin of these Raman peaks is substantiated by their absence in isostructural {\LP}, and their polarization dependence is in accordance with the lattice symmetry.
Raman spectroscopy thus lends itself as a viable laboratory-scale alternative for analyzing the crystal-field level structure of ultrathin metallic rare-earth compounds with high sensitivity and accuracy.

\section{acknowledgments}
We gratefully acknowledge financial support by the Deutsche Forschungsgemeinschaft through the research unit FOR 1162 (projects Ge1855/10-2 and Fa222/5-2) as well as experimental support by R. H{\"{o}}lldobler.


\providecommand{\noopsort}[1]{}\providecommand{\singleletter}[1]{#1}%

\end{document}